\documentclass[twocolumn,showpacs,preprintnumbers,amsmath,amssymb,prl,superscriptaddress]{revtex4-1}

\usepackage{amsfonts}
\usepackage[english]{babel}
\usepackage[T1]{fontenc}
\usepackage{times}
\usepackage{mathrsfs}
\usepackage{graphicx}
\usepackage{dcolumn}
\usepackage{bm}
\usepackage[colorlinks,bookmarks=false,citecolor=blue,linkcolor=red,urlcolor=blue]{hyperref}

\newcommand{\be}{\begin{eqnarray}}
\newcommand{\ee}{\end{eqnarray}}
\newcommand{\ba}{\begin{align}}
\newcommand{\ea}{\end{align}}

\usepackage{tabularx}

\usepackage[usenames,dvipsnames]{xcolor}
\usepackage{soul}
\setulcolor{BrickRed}
\setstcolor{purple}

\makeatletter
\newcommand*\dashline{\rotatebox[origin=c]{90}{$\dabar@\dabar@\dabar@$}}
\makeatother

\begin{document}
	\title{Topologically protected heat pumping from braiding Majorana zero modes}

	\author{Dganit Meidan}
	\affiliation{Department of Physics, Ben-Gurion University of the Negev, Beer-Sheva 84105, Israel}
	\author{Tal Gur}
	\affiliation{Department of Physics, Ben-Gurion University of the Negev, Beer-Sheva 84105, Israel}
	\author{Alessandro Romito}
	\affiliation{Department of Physics, Lancaster University, Lancaster LA1 4YB, United Kingdom}

\begin{abstract}
Majorana zero modes are non-Abelian quasiparticles that emerge on the edges of  topological phases of superconductors.
Evidence  of their presence  have been reported in transport measurements on engineered superconducting-based nanostructures. In this manuscript  we identify signatures of a topologically protected dynamical manipulation of Majorana zero modes via continuous transport measurement during the manipulation. Specifically, we show that, in a two-terminal geometry, the heat  pumped across the terminals at low temperature and voltage bias is characterized by a universal value. We show that this feature is inherent to the presence of Majorana zero modes and  discuss its robustness against temperature, voltage bias and the detailed coupling to the contacts. 

\end{abstract}
\maketitle
\vspace{.5pc}

\paragraph{Introduction}

Majorana zero modes are non-Abelian quasiparticles bound to the surface and to  defects  in topological supercoductors.\cite{Wilczek1982,Read1999,Stern2010} 
Driven by the promise of 
exploiting their non-Abelian statistics for fault-tolerant information processing \cite{Nayak2008}, proposals for engineering such a topological phase in superconductor-based 
nanostructures have shifted the focus on Majorana zero modes form a purely theoretical feature  \cite{Moore1991,Read1999,Kitaev2001} to an experimentally observable entity \cite{Fu2008,Lutchyn2010,Oreg2010}.  While  the first wave of  experiments \cite{Mourik2012,Das2012,Churchill2013,Nadj-Perge2014,Albrecht2016,Deng2016} were designed solely to detect these exotic excitations, the next generation of  experiments \cite{aasen2015,Plugge2016,Plugge2016a} and theoretical proposals \cite{Fu2010,Michaeli2016,Vijay2016,Rosenow2012,Rubbert2016,Romito2017} aim at probing directly their non-local  topological properties, with the ultimate goal of optimising and controlling their manipulation.

A key feature of topological phases based on superconducting nanostructures  
is the multitude  of available detection schemes. To date, transport properties, such as conductance and tunneling density of states \cite{Flensberg2010}, have provided the predominant  evidence of Majorana zero modes \cite{Mourik2012,Das2012,Churchill2013,Nadj-Perge2014,Albrecht2016,Deng2016}, and charge sensing schemes  have been proposed to effectively implement braiding protocols \cite{Bonderson2008,Bonderson2013a,Karzig2017}. Indeed,  having an open apparatus, where the topological material is coupled to non interacting  leads,  
allows to define new topological invariants based on scattering states\cite{Fulga2011,Meidan2014,Meidan2016}.
 
\begin{figure}
	\includegraphics[width=0.45\textwidth]{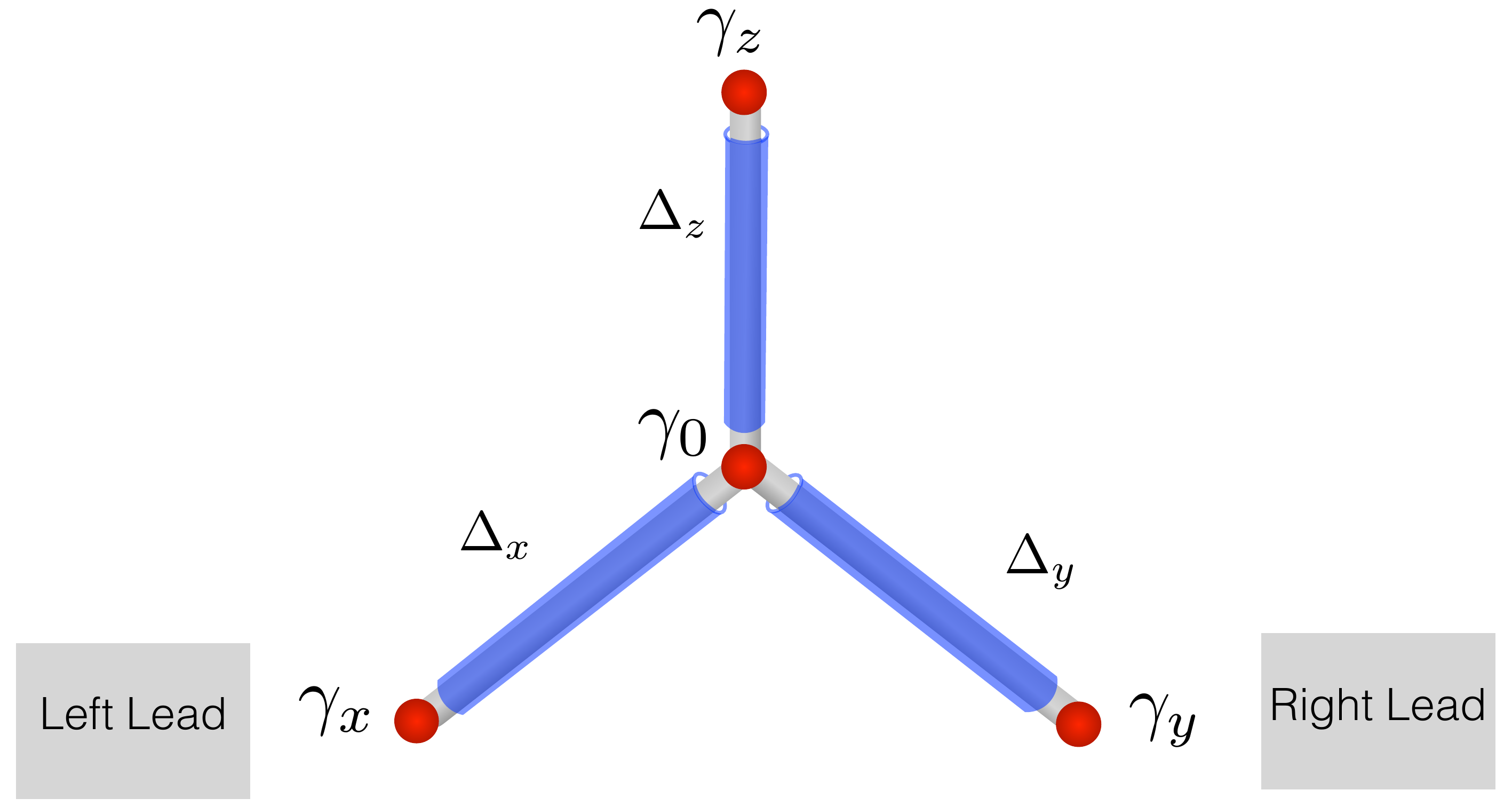} 
	\caption{Schematics of a junction of three 1d p-wave superconductors. The Majorana zero-modes at two ends of the junction are connected to external leads. The effective couplings between pairs of Majorana can be tuned independently.}
	\label{fig:system}
\end{figure}

\begin{figure}
\includegraphics[width=0.45\textwidth]{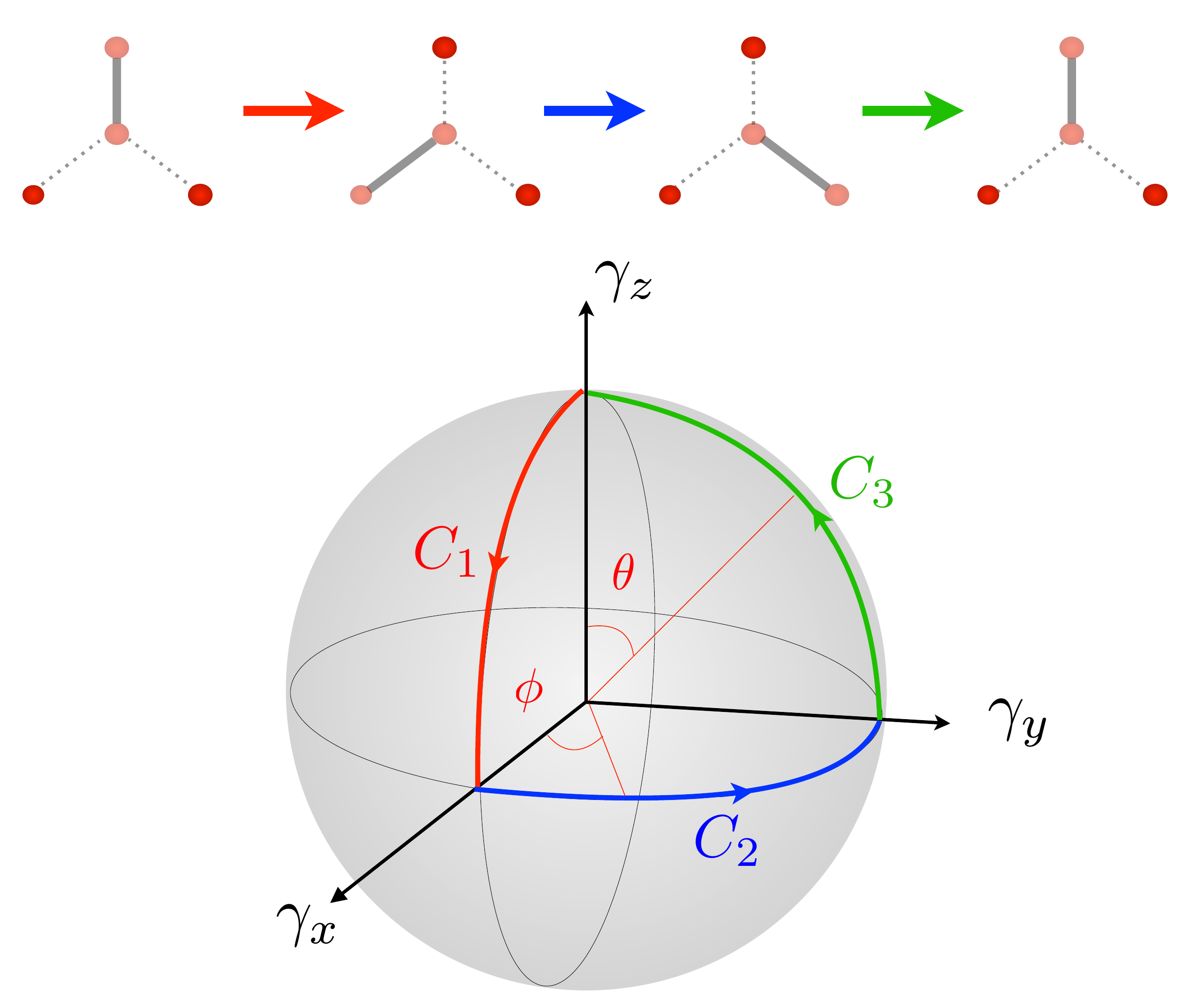}
	\caption{Schematic representation of the driving cycle. Grey lines represent the pairwise coupling between Majoranas ($\Delta_j \neq 0$ for full lines and $\Delta_j=0$ for dotted lines). The corresponding  stages of the evolution are represented as a path in the parameter space spanned by~$\vec{\Delta}$. }
	\label{fig:cycle_param}
\end{figure}

In this manuscript we look at an intermediate benchmark  towards the implementation and detection of  topologically protected operations  by examining the transport signatures of a braiding protocol.  
For this purpose we consider the minimal setup for the exchange of Majorana modes, in a Y-junction layout, see Fig. \ref{fig:system}. In a sequence of operations shown in Fig. \ref{fig:cycle_param}, the Y-junction setup is then driven to perform Majorana braiding. We show that  a combination of the density of  heat  and charge pumped during the braiding cycle is a universal quantity, independent of the details of the driving and of the couplings to the external leads.  We further discus how this measurable feature is affected by temperature and by the contact geometry.
This universal value stems directly from the geometric properties of the braiding cycle in parameter space, and can be used as a signature of such a topologically protected operation. It also provides a new potential tool for engineering heat pumping device.

\paragraph{Model ---}

The system we consider is schematically depicted in Fig.~\ref{fig:system}. It 
consists of  three topological superconducting (TSC) wires arranged in a Y-junction configuration. The wires can be realized in proximity coupled semi conducting wires with a strong spin orbit coupling, subjected to a magnetic field and proximity coupled to a superconductor \cite{Oreg2010,Lutchyn2010}. Throughout the manuscript we will assume that the superconductors are large enough such that charging effects can be neglected.
Each topological superconducting wire hosts two Majorana zero modes at its end. At energies well below the induced superconducting gap, $ \Delta$, the  Hilbert space is spanned by  three Majorana bound states    $\gamma_i$ where  $i={x,y,z}$ which appear at the three ends of the Y-junction, and a single Majorana bound state at the center $\gamma_0 $, as shown in the figure. 
(The junction may host additional  sub gap states. We restrict  to energies which are well below any accidental subgap excitations).
The four Majorana zero modes can be gapped in pairs, and the effective low energy Hamiltonian of the Y junction is  
\be
H_Y&=&i \gamma_0 \vec{\Delta}\cdot \vec{\gamma} 
\label{HY}
\ee
where $\vec{\Delta} = \Delta(\sin\theta\cos\phi,\sin\theta\sin\phi,\cos\theta) $ and $\vec{\gamma} = (\gamma_x,\gamma_y,\gamma_z) $.

In order to  study the transport  properties of the system we couple the Y-junction to two single channel leads, denoted by $L,R $. The full Hamiltonian is then given by $ H=H_Y+H_T+\sum_{\alpha=L,R} H_\alpha $, where 
\be\label{H}
H_T &=& t_L \left(c_{Lk}-c_{Lk}^\dag\right)\gamma_x+ t_R \left(c_{Rk}-c_{Rk}^\dag\right)\gamma_y\\
\nonumber
H_\alpha &=& \sum_k \xi_k c_{\alpha k}^\dag c_{\alpha k},
\ee
with $\xi_{\alpha k}$ the lead energy dispersion and with the tunneling charcterised by the rates $\Gamma_\alpha= \pi \nu_\alpha \vert t_\alpha \vert^2$, $\nu_\alpha$ the energy density of sates.

The Y-junction   is the simplest proposed setup  that allows to perform Majorana braiding. This can be achieved by adjusting the pairwise coupling $\Delta_i$  between in an adiabatic  sequence using gates or magnetic fluxes \cite{VanHeck2012,Karzig2016}.
The pairwise coupling sequence and the corresponding cycle in the parameter space are depicted in Fig.~\ref{fig:cycle_param}.
The induced operation on the state of the system  can be better understood by applying an orthogonal rotation on  $H_Y$ to obtain $ H_Y^d = i\Delta \gamma_0\gamma_r$,
where $\gamma_r = \vec{\gamma}\cdot  \hat{e}_r $, and the two zero modes $\gamma_\theta = \vec{\gamma}\cdot  \hat{e}_\theta $ and $\gamma_\phi= \vec{\gamma}\cdot  \hat{e}_\phi $, with $ \hat{e}_r$, $\hat{e}_\theta$ and $\hat{e}_\phi$ the basis vectors in spherical coordinates, and  $\hat{e}_\theta$ and $\hat{e}_\phi$ span a two fold degenerate subspace.  
The evolution induced by adjusting the pairwise coupling shown Fig.~\ref{fig:cycle_param} corresponds to adiabatically changing the projection of the Majorana bound states physically coupled to the leads $\gamma_{x,y,z}$ on to the degenerate subspace spanned by $\gamma_\theta,\gamma_\phi $.
Importantly the operation is exponentially protected regardless of the fluctuations in controlling the system parameters \cite{Karzig2016}.

\paragraph{Adiabatic change of scattering matrix --- }

The adiabatic manipulation of the system parameters depicted in Fig.~\ref{fig:cycle_param} can be regarded as pumping cycle when the system is coupled to external leads, see Fig.~\ref{fig:system}. In order to study the transport properties of the braiding cycle, we compute the  scattering matrix defined as $\Psi_{\rm out}(\epsilon) = S(\epsilon) \Psi_{\rm in}(\epsilon)$ where the scattering states $\Psi_{\rm in} = (\Phi_{e,L},\Phi_{h,L},\Phi_{e,R},\Phi_{h,R}) $ are in the basis of electrons $ (e)$ and holes $(h)$ in the left $(L) $ and right $(R)$ leads.
The scattering matrix can be expressed as \cite{Mahaux1969}:
\be\label{Smtx}
S(\epsilon) =  \openone +2 \pi i W^\dag \left( H_0-\epsilon-i\pi WW^\dag\right)^{-1} W
\ee
Where $W $ is the matrix that encodes the coupling to the leads. 
Eq. \eqref{Smtx} is more transparent in the rotated basis: $\tilde{\Psi} = (\Phi_{L}^+,\Phi_{R}^+,\Phi_{L}^-,\Phi_{R}^-) $ where $ \Phi_{\alpha}^\pm =  \frac{1}{\sqrt{2}}\left[\Phi_{e,\alpha}\pm \Phi_{h,\alpha}\right]$. From Eq \eqref{H} it follows that in this basis the scattering matrix takes a block diagonal form with only two channels coupled to the Y-junction. We therefore restrict to  the  two dimensional basis of coupled channels, namely $\psi=  (\Phi_{L}^-,\Phi_{R}^-)$.  Here the coupling matrix takes the simplified form $
W=\sqrt{2}t_L  |x\rangle \langle L^-| +\sqrt{2}t_R  |y\rangle \langle R^-| $, and the scattering matrix can be written as $S(\epsilon)=\mathbf{1}_2 \oplus \tilde{S}(\epsilon)$, where the $\tilde{S}(\epsilon) \in U(2)$ is the non-trivial block of the scattering matrix. 

\begin{figure}
	\includegraphics[width=0.45\textwidth]{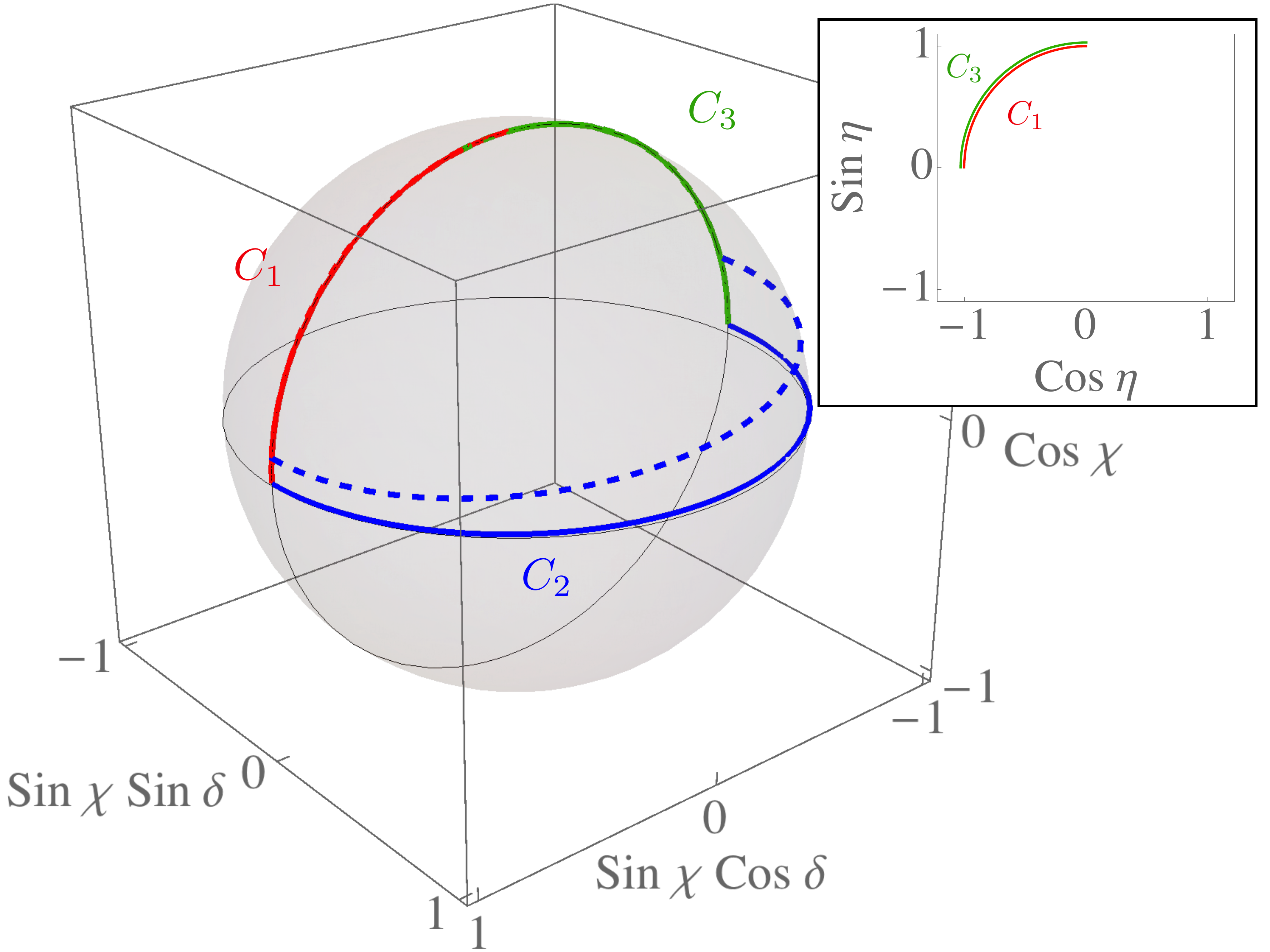} 	\caption{$\tilde{S}(\epsilon)$ along the braiding cycle. The color codes identifies the segments of the loop in parameter space depicted in Fig. \ref{fig:cycle_param}. Here $\Gamma_L/\Gamma_R=0.3$, $\epsilon/\Gamma_R=0.01$ (solid curve) and $\epsilon/\Gamma_R=0.1$ (dashed curve). The inset shows  a plot of  the $U(1)  $ phase of the scattering matrix, $\eta$,  for general  $\Gamma_L/\Gamma_R=0.3$ and $\epsilon/\Gamma_R=1\times10^{-3}. $ (the two curves are  shifted for clarity, the blue segment is not visible since $\eta$ is not affected)}.
	\label{fig:2LeadsGeneralCoupling}
\end{figure}

At energies well below the induced superconducting gap,  $ \epsilon \ll \Delta $, only the two fold degenerate ground-space of the Hamiltonian takes part in the transport, and we need to account for the lead-system coupling terms projected on the ground state only. Defining a projection operator $P_{\rm g} = |\theta\rangle \langle \theta| +|\phi\rangle \langle \phi|  $, we can rewrite the  expression for the scattering matrix in Eq. \eqref{Smtx} by replacing $W$ with its projected equivalent,
\be
\nonumber
P_{\rm g} W &=&\sqrt{2}t_L \left[  \cos\theta\cos\phi  |\theta\rangle \langle L^-|-\sin\phi   |\phi\rangle \langle L^-| \right]\\
&+&\sqrt{2}t_R   \left[ \cos\theta\sin\phi    |\theta\rangle \langle R^- |+\cos\phi  |\phi\rangle \langle R^-|  \right] .\label{eq:proiezione}
\ee
where $\langle \theta \vert x \rangle = \cos \theta \cos \phi$, $\langle \theta \vert y \rangle = \cos \theta \sin \phi$, $\langle \phi \vert x \rangle = -\sin \phi $, $\langle \phi \vert y \rangle = \cos \phi$.
Using Eq. \eqref{eq:proiezione} in Eq. \eqref{Smtx}, the resulting scattering matrix takes a simplified form:
\be
\nonumber
\tilde{S}(\epsilon)& =& e^{-i\eta(\epsilon) } e^{-i\chi(\epsilon) \left(\cos\delta(\epsilon)\sigma_z+\sin\delta(\epsilon)\sigma_x\right)} \\
&=&  e^{-i\eta } \left[\cos\chi -i\sin\chi \left( \cos\delta \sigma_z+\sin\delta \sigma_x\right)\right],  \label{eq:scattering}
\ee
where an analytical expression for $\eta(\epsilon)  \in [0, 2\pi)$, $\chi(\epsilon)  \in [0, \pi]$, and $\delta(\epsilon)  \in [0,  \pi )$  which depend on the parameters controlled along the cycle, $\theta$ and $\phi$, is given in the supplemental material \cite{suppl}.
For equal coupling strength $\Gamma_L=\Gamma_R$ the mapping takes a compact form: 
\be
\label{eq:mappa}
\nonumber
\tan \delta(\epsilon) &=& \tan 2\phi\\
\nonumber
\tan \chi(\epsilon) &=& \frac{\epsilon\Gamma\sin\theta^2}{\epsilon^2+\Gamma^2\cos\theta^2}\\
\tan \eta(\epsilon) &=& \frac{\epsilon\Gamma(1+\cos\theta^2)}{\epsilon^2-\Gamma^2\cos\theta^2}. 
\ee 
Eq. (\ref{eq:scattering}) maps the path in parameter space depicted in Fig. \ref{fig:cycle_param} to a path in the space of scattering matrices, as depicted  in Fig. \ref{fig:2LeadsGeneralCoupling}.  In the limit of low excitations energies  $\epsilon \ll \min{\Gamma_L,\Gamma_R}$ and for generic coupling strength $\Gamma_L\neq\Gamma_R $, the path $\mathcal{C}$ of $\tilde{S}(\epsilon)$ approaches the solid line in Fig. \ref{fig:2LeadsGeneralCoupling}.
 Note that  $\epsilon= 0$  gives a trivial path. This reflects the fact that the zero-energy limit conflicts with 
the exponentially small energy splitting due to the overlap of Majorana zero modes,  $\epsilon_0$, and a non trivial limit requires $\epsilon \gg \epsilon_0$.

\paragraph{Pumped charge and heat --- }

In order to identify a suitable transport property which reflects the global geometric properties of the path in parameter space (cf. Fig.\ref{fig:cycle_param}), we consider first the charge and heat current at lead $\alpha$, $I_{e,\alpha} =e/h \int d\epsilon [f_{\rm in} (\epsilon)- f_{\rm out}] (\epsilon)$ and  $I_{\epsilon,\alpha}= 1/h \int d\epsilon (\epsilon-\mu) [f_{\rm in} (\epsilon)- f_{\rm out}]$ respectively, where $f_{\rm in}(\epsilon)$ and $f_{\rm out} (\epsilon)$ are the ingoing and outgoing electronic energy distribution functions in the leads, kept at temperature $T$ and voltage bias $eV=\mu$. The latter is determined by the time dependent scattering matrix of the system. Following a well established procedure \cite{Moskalets2002}, the scattering matrix is expanded to include the leading small slow-varying corrections, to obtain the expression for the pumped charge and heat $\mathcal{Q}_{e(\epsilon),\alpha} =\int_0^\tau dt \, I_{e(\epsilon),\alpha}(t)$ in terms of the parametric dependence of the scattering matrix \cite{Brouwer1998}. The result are known for the case superconducting leads\cite{Moskalets2002,Blaauboer2002,Taddei2004}, and we extend them here to a normal-superconducting junction with a  Majorana zero modes -- see \cite{suppl}.

At finite bias $\mu\neq0 $, the braiding manipulation 
 gives rise to   heat and charge currents, both  accompanied by time-independent contributions.
However, at  $\mu=0$, as a direct consequence of particle-hole symmetry of the Majorana-lead coupling,  the charge current is {\it identically zero} even  in the presence of a driving.  That is not generally the case for the heat current, which embeds the geometrical contribution resulting from  the braiding protocol. In this limit, the heat pumping is given by $ \mathcal{Q}_{\epsilon,\alpha} (\mu=0, T) = \int_0^\tau dt \, I_{\epsilon,\alpha} = 2\int_0^\infty d\epsilon \, \epsilon \, \mathcal{Q}_{0,\alpha} (\epsilon) \partial f(\epsilon)/\partial \epsilon$ with 
\be
\label{eq:calore}
& &\mathcal{Q}_{0,\alpha}(\epsilon)=  \oint_C \left(\frac{dn(\alpha)}{d\theta}d\theta+\frac{dn(\alpha)}{d\phi}d\phi\right),
\ee
where  $f(\epsilon)$ is the Fermi distribution function and
\be
\nonumber
\frac{dn(\alpha)}{dX}&=&\frac{1}{ 2 \pi}  \sum_{\beta,\nu} {\rm Im}\frac{\partial  S^{e,\nu }_{\alpha,\beta}}{\partial X} {S^{e,\nu}_{\alpha,\beta}}^* \\ 
&=&\frac{1}{4\pi}  \sum_{r} {\rm Im} \frac{\partial  \tilde{S}_{\alpha^-,r}}{\partial X} \tilde{S}_{\alpha^-,r}^* -\frac{1}{4\pi} {\rm Im} \frac{\partial  \tilde{S}_{\alpha^-,\alpha^-}}{\partial X} \label{eq:tecnica}.
\ee
Here $\beta= L,R$, $\nu=e,h$ and $r= L^-,R^-$.
At low temperatures, $ T\to 0$, this expression reduces to:
\begin{equation}
\frac{\mathcal{Q}_{\epsilon,L}}{T} = 2 \ln 2 \, \lim_{\epsilon \to 0} {Q}_{0,L}(\epsilon) +\mathcal{O}(T).
\end{equation} 
The zero energy limit of $\mathcal{Q}_{0,\alpha}$ can be conveniently computed  using the mapping of the path in parameter space to that in the scattering matrix space, portrayed  in Fig.~\ref{fig:2LeadsGeneralCoupling}. 
In this limit, the heat pumped  approaches a universal value independent on the coupling parameters $\Gamma_L,\Gamma_R $. Form Eq. \eqref{eq:scattering} and Eq. (\ref{eq:calore}) we find: 
\be
\frac{\mathcal{Q}_{\epsilon,L}}{2 \ln 2 \, T}=\frac{1}{4} \label{eq:principale}
\ee

 \begin{figure}
	\includegraphics[width=0.45\textwidth]{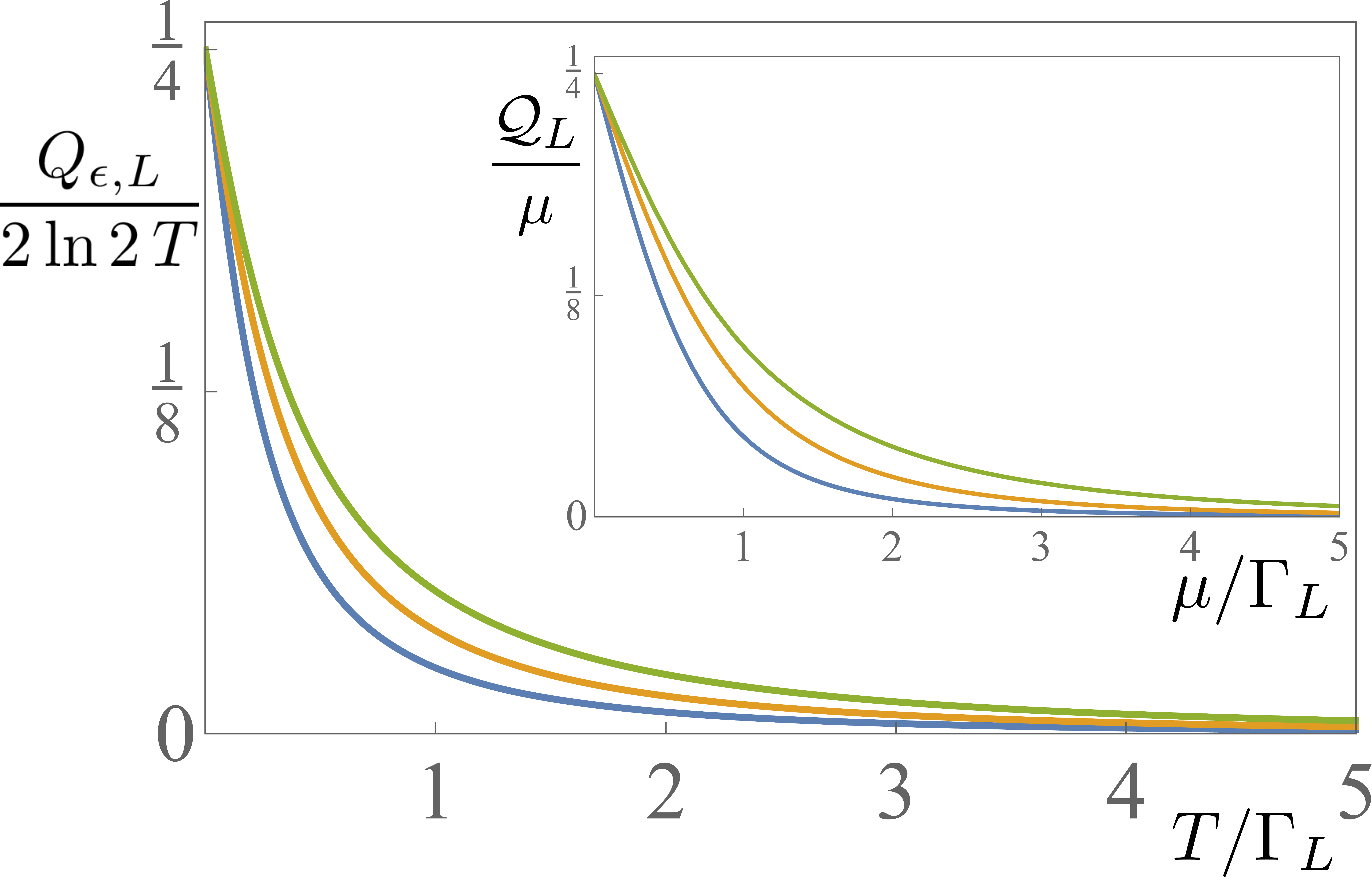} 	\caption{The heat pumped into the left lead during a braiding process, $\frac{\mathcal{Q}_{\epsilon,L}}{2 \ln 2 \, T}$, as a function of the temperature $ T/\Gamma_L$. Different curves correspond to different values of the coupling $ \Gamma_R/\Gamma_L = 1$ (blue curve), $ \Gamma_R/\Gamma_L = 2$ (yellow curve), $ \Gamma_R/\Gamma_L = 4$ (Green curve). Inset shows the combined current and heat pumped during a braiding process as a function of the chemical potential $ \mu/\Gamma_L$ at zero temperature, for different values of the coupling constants (same color code).}\label{fig:QL2}
\end{figure}
Eq. (\ref{eq:principale}) is the main result of our paper. A few comments are in place. We first note that the braiding protocol defines a path in parameter space (see Fig. \ref{fig:cycle_param}). 
At low scattering energies $\epsilon \to 0$ this path is mapped onto a distinctive path in the scattering matrix space, independent on the coupling to leads, which in turns leads to the universal  value of  $\mathcal{Q}_{\epsilon,\alpha}/T$. This universal value   is therefore inherited from the geometrical properties of the path in parameter space,  and reflects its topological protection, i.e. the measured heat pumped is protected against fluctuations of the coupling and the cycle driving.
Second, in the absence of Majorana zero energy excitations, the ground state degeneracy is lifted. This corresponds to setting $\epsilon=0$ in Eqs. (\ref{eq:scattering},\ref{eq:mappa}), where the path in parameter space would be trivial and there would be no pumped charge or heat. Finally, we see from Eq. \eqref{eq:tecnica} that $\mathcal{Q}_{\epsilon,R}=-\mathcal{Q}_{\epsilon,L}$. This anti-correlation of pumped  heat  at the two leads is distinctive of the Majorana braiding cycle, and can be used to distinguish the presence of Majorana bound states from a local zero energy  Andreev bound state, which would generically result in {\it uncorrelated}  heat transfer at  the two leads.

The value of the pumped heat $\mathcal{Q}_{\epsilon,\alpha}$, is plotted as a function of $T$ in Fig. \ref{fig:QL2}. The universal value $\mathcal{Q}_{\epsilon,\alpha}/2 \ln 2 \, T=1/4 $ is obtained under the physical conditions $T \ll \mu \ll \min\{\Gamma_L,\Gamma_R \}$. Away from this  limit, at finite $T$,  the pumped heat 
deviates form the universal value in a parameter dependent fashion. 
Applying a finite bias $\mu $, introduces time-independent contributions to the charge and heat currents which originate from Andreev reflection processes, present also in the absence of the adiabatic manipulation. In this case,  the combination that singles out the  contribution  from the braiding process is $ \mathcal{Q}_\alpha \equiv \mathcal{Q}_{\epsilon,\alpha} + \frac{\mu}{e}\mathcal{Q}_{e,\alpha}$ expressed by:
\be
& & \mathcal{Q}_{\alpha} = \int_0^\infty d\epsilon \, \epsilon \, \mathcal{Q}_{0,\alpha} (\epsilon) \left[  \frac{\partial f(\epsilon-\mu)}{\partial \epsilon} + \frac{\partial f(\epsilon+\mu)}{\partial \epsilon} \right] ,
\ee
with $\mathcal{Q}_{0,\alpha} $ given by Eq. \eqref{eq:calore}.
As long as $\mu \ll T$, a finite bias induces small corrections of the universal value. When $T \ll \mu$, instead, we have a different picture  and a finite value of $\mathcal{Q}_\alpha$ is induced by the finite chemical potential. 
In the limit $T =0$, and $\mu \to 0$ we have
\be
\frac{\mathcal{Q}_L}{\mu} =  \lim_{\epsilon \to 0} \mathcal{Q}_{0,L}(\epsilon) = \frac{1}{4}
\ee 
Here the role of temperature is replaced by the chemical potential. The non abelian phase acquired during the pumping cycle is embedded in the linear dependence on the pumped heat on chemical potential. The general dependence on the bias $\mu$ is shown  in the inset of Fig. \ref{fig:QL2}.
As in the limit of $\mu = 0$ and finite temperature, the deviation from the universal value occurs at $\mu \sim \min\{\Gamma_L, \Gamma_R\}$.

\paragraph{Three-terminal geometry --- }
\begin{figure}
	\includegraphics[width=0.45\textwidth]{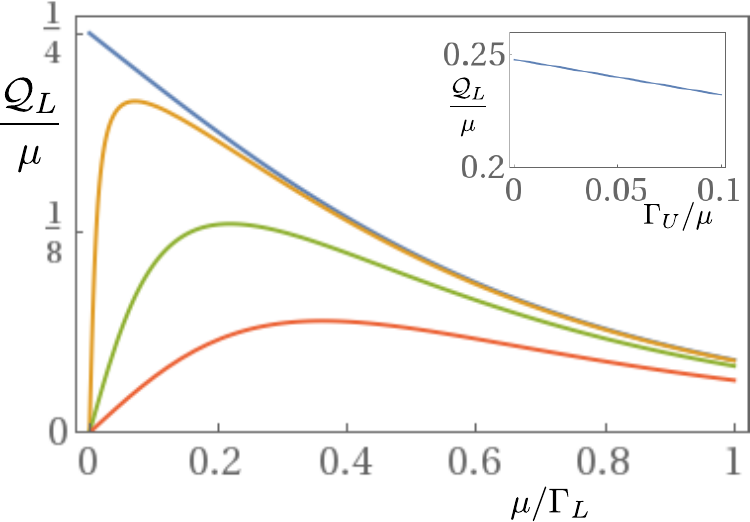}
	\caption{The differential heat per energy pumped into the left lead during a braiding process in the three lead configuration. Plots where calculated for $\Gamma_L/\Gamma_R =1 $ and different  coupling of the third lead $ \Gamma_U/\Gamma_L = 0$ (blue curve), $ \Gamma_U/\Gamma_L = 0.01$ (yellow curve), $ \Gamma_U/\Gamma_L = 0.1$ (Green curve) and $ \Gamma_U/\Gamma_L = 0.3$ (red curve). Inset shows the differential heat pumped into the left lead as a function of the ratio $ \Gamma_U/\mu$. Here $ \Gamma_L/\Gamma_R = 0.3$ and $\mu/\Gamma_L =0.01$.}
	\label{fig:QL3}
\end{figure}
We next  study the case where the three arms of the Y-junction setup are coupled to leads. This is described by the Hamiltonian $H= H_Y+H_{3T}+\sum_{\alpha} H_\alpha$ with the same notation as in the two-leads case  and where $ H_{3T} =
H_T+ t_U \left(c_{Uk}-c_{Uk}^\dag\right)\gamma_z
$
and $\alpha = L,R,U $. Following the steps outlined above, we calculate the scattering matrix using Eq. \eqref{Smtx}. In the rotated basis: $\tilde{\Psi} = (\Phi_{L}^+,\Phi_{R}^+,\Phi_{U}^+,\Phi_{L}^-,\Phi_{R}^-,\Phi_{U}^-) $ where $ \Phi_{\alpha}^\pm =  \frac{1}{\sqrt{2}}\left[\Phi_{e,\alpha}\pm \Phi_{h,\alpha}\right]$, the scattering matrix takes a block diagonal form, $S(\epsilon) =\mathbf{1}_3 \oplus \tilde{S}(\epsilon)$, with $\tilde{S}(\epsilon) \in U(3)$. Restricting to  the  three dimensional basis of coupled channels, $\psi=  (\Phi_{L}^-,\Phi_{R}^-,\Phi_{U}^-)$ and projecting  onto the ground state manifold, the coupling matrix takes the form $
 P_{\rm g} W 
 =\sqrt{2}t_L\left[  \cos\theta\cos\phi  |\theta\rangle \langle L^-|-\sin\phi   |\phi\rangle \langle L^-| \right]
+\sqrt{2}t_R   \left[ \cos\theta\sin\phi    |\theta\rangle \langle R^- |+\cos\phi  |\phi\rangle \langle R^-|  \right] + \sqrt{2}t_U\left[  \sin\theta |\theta\rangle \langle U^-| \right]$

In the three terminal geometry, $\tilde{S}(\epsilon)$ takes a more complicated form than that in Eq. (\ref{eq:scattering})(cf. Ref. [\onlinecite{suppl}] for details). 
The combined  heat and charge pumped along the cycle is then calculated using   Eq. \eqref{eq:calore}.
Fig. \ref{fig:QL3} depicts the combined  heat and charge, $Q_L/\mu$ pumped  during a braiding process for different values of the coupling strength to the upper lead. Interestingly, the universal value at vanishing energy is lost once the broadening due to the coupling to the third lead is of the order of $\mu$. Therefore $\mathcal{Q}_{L}/\mu$ approaches $1/4$ only in the regime $\Gamma_U \ll \mu \ll \min{\Gamma_R,\Gamma_L}$.
This result is to be expected, as the limit $\mu \ll \min{\Gamma_U, \Gamma_L, \Gamma_R}$  is adiabatically connected to the limit of equal coupling, where we expect no charge transfer as a result of symmetry.

\paragraph{Conclusions--}

In conclusion, we have analysed a quantum device in which Majorana zero modes undergo a dynamical braiding operation, while the system is continuously  probed by external leads. Considering explicitly the implementation in superconducting nanostructures where two Majorana end-states are contacted to external leads, we have related the accumulated non-abelian geometric phase to a surface integral in the space of scattering matrices. In  the limit of low energy,  this results in  a universal value of the  heat density 
pumped during a cycle. (Notably, the pumped charge is identically zero).  We have analysed the robustness of the effect against the variation of parameters, temperature, and setup geometry. Our results establishes a first connection between transport properties and topologically protected dynamical operations in Majorana based devices. 
 The identified universal property can provide a new ingredient for engineering heat pumping devices and for understanding the role of topological protection in thermodynamic processes.

\paragraph{Acknowledgements--}
We acknowledge discussions with Piet Brouwer, Dimitri Gutman, Tami Pereg-Barnea, and Janine Splettst\"osser.
A.R. acknowledges support by EPSRC via Grant No.
EP/P010180/1. 
D. M. acknowledges support from the Israel Science Foundation (Grant No. 737/14) and from the People Programme (Marie Curie Actions) of the European Union's Seventh Framework Programme (FP7/2007-2013) under REA grant agreement No. 631064. 


\newpage

\onecolumngrid
\part{Supplementary Materials}

\setcounter{equation}{0}
\setcounter{figure}{0}
\setcounter{table}{0}
\setcounter{page}{1}
\makeatletter
\renewcommand{\theequation}{S\arabic{equation}}
\renewcommand{\thefigure}{S\arabic{figure}}
\renewcommand{\bibnumfmt}[1]{[S#1]}
\renewcommand{\citenumfont}[1]{S#1}

\pagestyle{empty}

Here we present the derivation of the general form of the scattering matrix for the geometry with 2 and 3 contact leads. We  derive a general formulas for pumped heat and charge for a system consisting of both superconducting and normal leads and apply it to the device considered in the paper.

\section{General form of the scattering matrix for 2 and 3 leads}

We study first the case where we couple the superconducting Island to 3 leads. The setup is described by the Hamiltonian:
\be
\nonumber
H_0&=&2i \gamma_0 \vec{\Delta}\cdot \vec{\gamma} \\
\nonumber\label{H3}
H_T &=& t_L \left(c_{Lk}-c_{Lk}^\dag\right)\gamma_x+t_R \left(c_{Rk}-c_{Rk}^\dag\right)\gamma_y + t_U \left(c_{Uk}-c_{Uk}^\dag\right)\gamma_z\\
\nonumber
H_\alpha &=& \sum_k \xi_k c_{\alpha k}^\dag c_{\alpha k}
\ee
Analogously to the procedure in the paper, we calculate the unitary scattering matrix: $\Psi_{\rm out}(\epsilon) = S(\epsilon) \Psi_{\rm in}(\epsilon)$ where the scattering states $\Psi_{\rm in} = (\Phi_{e,L},\Phi_{h,L},\Phi_{e,R},\Phi_{h,R},\Phi_{e,U},\Phi_{h,U}) $ are in the basis of electrons $ (e)$ and holes $(h)$ in the left and right leads. 
The scattering matrix can be expressed as:
\be
S(\epsilon) =  \openone +2 \pi i W^\dag \left( H_0-\epsilon-i\pi WW^\dag\right)^{-1} W
\ee
Where $W $ is the matrix that encodes the coupling to the leads. It is possible to perform a unitary transformation that decouples some of the leads degrees of freedom form the system. We work in this rotated basis defined by  $\tilde{\Psi} = (\Phi_{L}^+,\Phi_{R}^+,\Phi_{U}^+,\Phi_{L}^-,\Phi_{R}^-,\Phi_{U}^-) $ where $ \Phi_{\alpha}^\pm =  \frac{1}{\sqrt{2}}\left[\Phi_{e,\alpha}\pm \Phi_{h,\alpha}\right]$. It follows directly that From Eq. \eqref{H3} that, in this basis, the scattering matrix takes a block diagonal form with only three channels coupled to the dot. We therefore restrict to  the  three dimensional basis of coupled channels, namely $\psi=  (\Phi_{L}^-,\Phi_{R}^-,\Phi_{U}^-)$ . 
Denoting  $\Gamma_U$ the coupling to the third lead, the coupling matrix takes the form:
\be
W &=& \left(
\begin{array}{cccc}
	\sqrt{\Gamma_L}&0&0\\
	0&\sqrt{\Gamma_R}&0 \\
	0&0&\sqrt{\Gamma_U}     \\
	0&0&0
\end{array}
\right) =\sqrt{\Gamma_L}  |x\rangle \langle L^-| +\sqrt{\Gamma_R}|y\rangle \langle R^-| +\sqrt{\Gamma_U}|z\rangle \langle U^-|
\ee
Projecting  onto the ground state manifold: $P_{\rm g} = |\theta\rangle \langle \theta| +|\phi\rangle \langle \phi|  $. We can  write:
\be
\nonumber
P_{\rm g} W &=&\sqrt{\Gamma_L} \left[ \langle \theta |x\rangle    |\theta\rangle \langle L^-|+\langle \phi |x\rangle    |\phi\rangle \langle L^-| \right]
+ \sqrt{\Gamma_R} \left[ \langle \theta |y\rangle    |\theta\rangle \langle R^- |+\langle \phi |y\rangle    |\phi\rangle \langle R^-|  \right]  + \sqrt{\Gamma_U} \left[ \langle \theta |z\rangle    |\theta\rangle \langle U^-|+\langle \phi |z\rangle    |\phi\rangle \langle U^-| \right]\\
&=&\sqrt{\Gamma_L}\left[  \cos\theta\cos\phi  |\theta\rangle \langle L^-|-\sin\phi   |\phi\rangle \langle L^-| \right]+ \sqrt{\Gamma_R}   \left[ \cos\theta\sin\phi    |\theta\rangle \langle R^- |+\cos\phi  |\phi\rangle \langle R^-|  \right] - \sqrt{\Gamma_U}\left[  \sin\theta |\theta\rangle \langle U^-| \right]
\ee
The non-trivial part of the scattering matrix follows:
\be
\label{eq:3contatti}
\nonumber
\tilde{S}(\epsilon) &=&  \openone +2  i W^\dag \left( P_{\rm g}[H_0-\epsilon-iWW^\dag] P_{\rm g} \right)^{-1} W\\
\nonumber
&=& \openone -2  i W^\dag \left( \epsilon+iP_{\rm g}WW^\dag P_{\rm g} \right)^{-1} W\\
\nonumber
&=& |L^-\rangle \langle L^-|\left\{ 1- 2i \Gamma_L \left[\frac{\left(\epsilon+i \Gamma_R \right)\cos^2\theta+\left(\epsilon+i \Gamma_U \right)\sin^2\theta\sin^2\phi}{\left(\epsilon+i \Gamma_U \right)\sin^2\theta\left(\epsilon+i \Gamma_L\sin^2\phi+i \Gamma_R\cos^2\phi \right)+\left(\epsilon+i \Gamma_L \right)\left(\epsilon+i \Gamma_R \right)\cos^2\theta}\right]\right\}\\
\nonumber
&+& |R^-\rangle \langle R^-|\left\{ 1- 2i \Gamma_R\left[\frac{\left(\epsilon+i \Gamma_L \right)\cos^2\theta+\left(\epsilon+i \Gamma_U \right)\sin^2\theta\cos^2\phi}{\left(\epsilon+i \Gamma_U \right)\sin^2\theta\left(\epsilon+i \Gamma_L\sin^2\phi+i \Gamma_R\cos^2\phi \right)+\left(\epsilon+i \Gamma_L \right)\left(\epsilon+i \Gamma_R \right)\cos^2\theta}\right]\right\}\\
\nonumber
&+& |U^-\rangle \langle U^-|\left\{ 1- 2i \Gamma_U\sin^2\theta\left[\frac{\left(\epsilon+i \Gamma_L\sin^2\phi+i \Gamma_R\cos^2\phi \right)}{\left(\epsilon+i \Gamma_U \right)\sin^2\theta\left(\epsilon+i \Gamma_L\sin^2\phi+i \Gamma_R\cos^2\phi \right)+\left(\epsilon+i \Gamma_L \right)\left(\epsilon+i \Gamma_R \right)\cos^2\theta}\right]\right\}\\
\nonumber
&-& 2i\sqrt{\Gamma_L\Gamma_R} \left[|L^-\rangle \langle R^-|+|R^-\rangle \langle L^-|\right]\left\{\frac{\left(\epsilon+i \Gamma_U \right)\sin^2\theta\cos\phi\sin\phi}{\left(\epsilon+i \Gamma_U \right)\sin^2\theta\left(\epsilon+i \Gamma_L\sin^2\phi+i \Gamma_R\cos^2\phi \right)+\left(\epsilon+i \Gamma_L \right)\left(\epsilon+i \Gamma_R \right)\cos^2\theta}\right\}\\
\nonumber
&+& 2i\sqrt{\Gamma_U\Gamma_R}\left[ |U^-\rangle \langle R^-|+ |R^-\rangle \langle U^-|\right]\left\{\frac{\left(\epsilon+i \Gamma_L \right)\cos\phi\cos\theta\sin\theta}{\left(\epsilon+i \Gamma_U \right)\sin^2\theta\left(\epsilon+i \Gamma_L\sin^2\phi+i \Gamma_R\cos^2\phi \right)+\left(\epsilon+i \Gamma_L \right)\left(\epsilon+i \Gamma_R \right)\cos^2\theta}\right\}\\
&+& 2i\sqrt{\Gamma_U\Gamma_L}\left[ |U^-\rangle \langle L^-|+|L^-\rangle \langle U^-|\right]\left\{\frac{\left(\epsilon+i \Gamma_R \right)\sin\phi\cos\theta\sin\theta}{\left(\epsilon+i \Gamma_U \right)\sin^2\theta\left(\epsilon+i \Gamma_L\sin^2\phi+i \Gamma_R\cos^2\phi \right)+\left(\epsilon+i \Gamma_L \right)\left(\epsilon+i \Gamma_R \right)\cos^2\theta}\right\}\
\ee

\begin{figure}
	\includegraphics[width=0.45\textwidth]{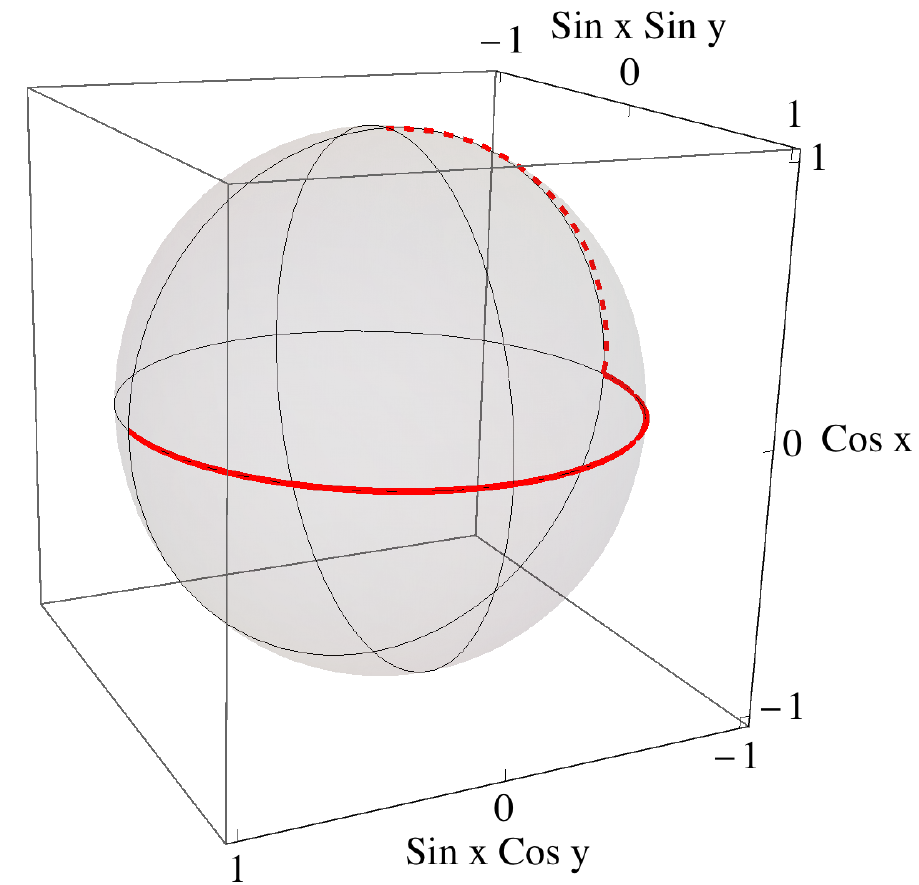}
	\caption{The evolution  the $x $ and $y $ angles of the scattering matrix, Eq. \eqref{Smtx3C1-1} during the first segment of the pumping cycle.  The solid line was calculated for $\Gamma_U/\Gamma_L= 0.3$ and $\epsilon/\Gamma_L= 10^{-4} $, while the dashed line shows the limit of  uncoupling the upper lead$ \Gamma_L\gg \epsilon\gg\Gamma_U$, where $\epsilon/\Gamma_L= 10^{-2}$ and $\Gamma_U/\Gamma_L= 10^{-5}$.}
	\label{fig:3LeadsGeneralCoupling}
\end{figure}
The expression in Eq. \eqref{eq:3contatti} simplifies considerably when the scattering matrix is evaluated along the path in parameter space of the pumping cycle. We obtain, for the three branches of the cycle:
\be\label{Smtx3C1-1}
\nonumber
S(\mathcal{C}_1) &=& e^{-ix_{\vec{C}_1}}\left[ \cos x_{\vec{C}_1} \frac{4 \mathbf{1} +3 \lambda_3 -\sqrt{3}\lambda_8}{6}
+i\sin x_{\vec{C}_1}\cos y_{\vec{C}_1}\frac{\sqrt{3}\lambda_8+\lambda_3}{2}
+i\sin x_{\vec{C}_1}\sin y_{\vec{C}_1}\lambda_4
\right]
+ e^{-2i\zeta_R} \frac{2 \mathbf{1} -3 \lambda_3 +\sqrt{3}\lambda_8}{6}
\\
\nonumber
S(\mathcal{C}_2) &=& e^{-ix_{\vec{C}_2}}\left[ \cos x_{\vec{C}_2} \frac{2 \mathbf{1} +\sqrt{3}\lambda_8}{3}
+i\sin x_{\vec{C}_2}\cos y_{\vec{C}_2}\lambda_3
+i\sin x_{\vec{C}_2}\sin y_{\vec{C}_2}\lambda_1
\right] + e^{-2i\zeta_U} \frac{\mathbf{1} -\sqrt{3}\lambda_8}{3}
\nonumber \\
S(\mathcal{C}_3) &=& e^{-ix_{\vec{C}_3}}\left[ \cos x_{\vec{C}_3} \frac{4 \mathbf{1} -3 \lambda_3 -\sqrt{3} \lambda_8}{6}
+i\sin x_{\vec{C}_3}\cos y_{\vec{C}_3}\frac{1}{2}\left[\sqrt{3}\lambda_8-\lambda_3\right]
+i\sin x_{\vec{C}_3}\sin y_{\vec{C}_3}\lambda_6
\right]+ e^{-2i\zeta_L} \frac{2 \mathbf{1} +3 \lambda_3 +\sqrt{3} \lambda_8}{6}
\nonumber
\ee
where
$\vec{C}_1= (L,U,\theta)$, $\vec{C}_2=(R,L,\phi)$, $\vec{C}_3= \vec{C}_3=(U,R,\phi)$ and 
$\tan x_{(\alpha,\beta,\psi)} = (\Gamma_\alpha\cos\psi^2+\Gamma_\beta\sin\psi^2)/\epsilon$,
$\tan y_{(\alpha,\beta,\psi)} =(2\sqrt{\Gamma_\alpha\Gamma_\beta}\cos\psi\sin\psi)/(\Gamma_\beta\sin\psi^2-\Gamma_\alpha\cos\psi^2)$
and $\tan\zeta_{\alpha} =\Gamma_\alpha/\epsilon$.
The scattering matrix dependence on the parameters is represented graphically in Fig. \ref{fig:3LeadsGeneralCoupling}, which presents the $x$ and $y$ components along the path in parameter space.

\section{General form of the Scattering matrix for 2 leads}
Upon decoupling one of the leads, by setting $ \Gamma_U =0$ in  Eq. \eqref{eq:3contatti}, the scattering matrix reduces to that of the two-leads geometry, $\tilde{S}(\epsilon) \in SU(2)$, which can be parametrised in terms of scattering matrices as
\label{eq:s2conatti1}
\be
\nonumber
\tilde{S}(\epsilon) = e^{-i\eta(\epsilon) } e^{-i\chi(\epsilon) \left(\cos\delta(\epsilon)\sigma_z+\sin\delta(\epsilon)\sigma_x\right)} = e^{-i\eta(\epsilon) } \left[\cos\chi(\epsilon)  -i\sin\chi(\epsilon) \left( \cos\delta(\epsilon) \sigma_z+\sin\delta(\epsilon) \sigma_x\right)\right]
\ee
with
\be
\nonumber
\tan \delta(\epsilon)&=& \frac{2\sqrt{\Gamma_L\Gamma_R}\sin\theta^2\sin 2\phi}{(\Gamma_L-\Gamma_R)(1+\cos\theta^2)-(\Gamma_L+\Gamma_R)(\sin\theta^2\cos2\phi)}\xrightarrow[]{\Gamma_L\rightarrow \Gamma_R}  \tan 2\phi\\
\nonumber
\tan\chi(\epsilon) &=&\frac{\epsilon\sqrt{\left[(\Gamma_L-\Gamma_R)(1+\cos\theta^2)-(\Gamma_L+\Gamma_R)(\sin\theta^2\cos2\phi)\right]^2+4\Gamma_L\Gamma_R\sin\theta^4\sin 2\phi^2}}{2\left[\epsilon^2+\Gamma_L\Gamma_R\cos\theta^2\right]}\xrightarrow[]{\Gamma_L\rightarrow \Gamma_R}  \frac{\epsilon\Gamma\sin\theta^2}{\epsilon^2+\Gamma^2\cos\theta^2}\\
\tan\eta(\epsilon) &=&\frac{ \epsilon \left[\Gamma_L(\cos\theta^2+\sin\theta^2\sin\phi^2)+\Gamma_R(\cos\theta^2+\sin\theta^2\cos\phi^2)\right]}{\epsilon^2-\Gamma_L\Gamma_R\cos\theta^2}\xrightarrow[]{\Gamma_L\rightarrow \Gamma_R}\frac{\epsilon\Gamma(1+\cos\theta^2)}{\epsilon^2-\Gamma^2\cos\theta^2}
\ee
In the limit of equal coupling we recover the simplified expressions (6) in the paper.

\section{Pumped charge and heat with normal and superconducting leads}

The main result of the paper establishes a relation between the geometric phases associated with the braiding of Majorana zero modes and transport across the system. Specifically the pumped heat across the cycle encodes informations on the geometric phase.
We derive here the general expression for the pumped charge and heat in setups consisting of both normal and superconducting leads, and use it for the Majorana-based device studied in the paper.

The starting point for our derivation are the results obtained for charge pumping \cite{Moskalets2002,Blaauboer2002,Taddei2004}. The system under consideration consists of $N$ normal leads contacted to a mesocopic device in the presence of a grounded superconductor. The amplitude for electrons and holes injected from lead $\alpha$ to be reflected /transmitted as electrons or holes at lead $\beta$ is determined by the scattering matrix $S_{\beta,\alpha}^{(ij)}(t)$, where $i,j \in \{ e,h \}$, with the same notation used in the paper. 
The outgoing charge and energy currents at lead $\alpha$ are given by
\be
I_{e,\alpha} & = &\frac{e}{h}\int_0^\infty dE \left[ f_{{\rm out},\alpha}(E) -f_{{\rm in},\alpha} (E) \right], \\
I_{\epsilon,\alpha} & = &\frac{1}{h} \int_0^\infty dE (E-\mu_\alpha) \left[ f_{{\rm out},\alpha}(E) -f_{{\rm in},\alpha} (E) \right], 
\ee
Here the distribution of the ingoing particles at lead $\alpha$ is the equilibrium distribution set by the external temperature and chemical potential, $f_{{\rm in},\alpha} (E) = f_0(E-\mu_\alpha)$, and the outgoing one is determined by the scattering properties of the systems. 

In the following we simplify $\left[ f_{{\rm out},\alpha}(E) -f_{{\rm in},\alpha} (E) \right]$, and the result can be used for both the charge and heat current.
We assume that all the leads are kept at the same chemical potential, $\mu_\alpha=\mu$.
Moreover, the time dependence of the scattering matrix on time is due to the weak slow periodic driving of external parameters, $X_j(t)=X_j+X_{j,\omega} e^{i(\omega t-\phi_j)} + X_{j,\omega} e^{-i(\omega t-\phi_j)}$. The time-dependence of the scattering matrix is then expressed given by
\begin{equation}
\hat{S}(t)= \hat{S} + \hat{S}_{\omega}e^{-i\omega t} +\hat{S}_{-\omega} e^{i \omega t},
\end{equation}
with
\begin{equation}
\label{eq:scatter-time}
\hat{S}_{\pm \omega} = \sum_j X_{j,\omega}e^{\mp i \phi_j} \frac{\partial \hat{S}}{\partial X_j}.
\end{equation}
This essentially corresponds to including the scattering process involving the nearest energy side-bands. The distribution of outgoing electrons at energy $E$ is therefore determined by the distribution of ingoing electrons and holes at energies $E$, $E+\omega$ and $E-\omega$.
With the observation that the ingoing equilibrium distribution of holes is given by $f^{(h)}_{{\rm in}, \beta}(E) = f_0(E +\mu)$,
we get
\be
\label{eq:distribuzioni}
f_{{\rm out}, \alpha} (E) = \sum_{\beta}& & \left[ |S_{\alpha,\beta}^{ee}|^2 f_0 (E-\mu) + |S_{-\omega,\alpha,\beta}^{ee}|^2 f_0 (E-\mu+\omega) + |S_{\omega,\alpha,\beta}^{ee}|^2 f_0 (E-\mu-\omega) \right. \nonumber \\
& & \left. +|S_{\alpha,\beta}^{eh}|^2 f_0 (E+\mu) + |S_{-\omega,\alpha,\beta}^{eh}|^2 f_0 (E+\mu+\omega) + |S_{\omega,\alpha,\beta}^{eh}|^2 f_0 (E+\mu-\omega)
\right].
\ee
Expanding Eq. \eqref{eq:distribuzioni} at leading order in $\omega \to 0$, and using the unitarity of the scattering matrix, $\left( \hat{S} + \hat{S}_{\omega}e^{-i\omega t} +\hat{S}_{-\omega} e^{i \omega t}\right) \cdot \left( \hat{S}^\dagger + \hat{S}^\dagger_{\omega}e^{+i\omega t} +\hat{S}^\dagger_{-\omega} e^{-i \omega t}\right)= 1 $, we get 
\be
\label{eq:correnti}
I_{e,\alpha} &=& \frac{e}{h} \int_{-\infty}^{\infty} dE \left[ \left( f_0(E+\mu) -f_0(E-\mu ) \right) \left(1-|S_{\alpha,\beta}^{ee}|^2 -|S_{\omega,\alpha,\beta}^{ee}|^2 -|S_{-\omega,\alpha,\beta}^{ee}|^2 \right) \phantom{\frac{a}{b}} \right. \nonumber \\
& & \left.  + \hbar \omega \frac{\partial f (E-\mu)}{\partial E} \left( |S_{-\omega, \alpha,\beta}^{ee}|^2 - |S_{\omega,\alpha,\beta}^{ee}|^2 \right)
+ \hbar \omega \frac{\partial f (E+\mu)}{\partial E} \left( |S_{-\omega, \alpha,\beta}^{eh}|^2 - |S_{\omega,\alpha,\beta}^{eh}|^2 \right), \right]
\ee
and
\be
\label{eq:correnti2}
I_{\epsilon,\alpha} &=&  \frac{1}{h}  \int_{-\infty}^{\infty} dE (E-\mu) \left[ \left( f_0(E+\mu) -f_0(E-\mu ) \right) \left(1-|S_{\alpha,\beta}^{ee}|^2 -|S_{\omega,\alpha,\beta}^{ee}|^2 -|S_{-\omega,\alpha,\beta}^{ee}|^2 \right) \phantom{\frac{a}{b}} \right. \nonumber \\
& & \left.  + \hbar \omega \frac{\partial f (E-\mu)}{\partial E} \left( |S_{-\omega, \alpha,\beta}^{ee}|^2 - |S_{\omega,\alpha,\beta}^{ee}|^2 \right)
+ \hbar \omega \frac{\partial f (E+\mu)}{\partial E} \left( |S_{-\omega, \alpha,\beta}^{eh}|^2 - |S_{\omega,\alpha,\beta}^{eh}|^2 \right) \right],
\ee
where we have shifted the energy to be measured from the superconductor chemical potential and we have neglected the effect of finite band size in extending the integral over all energies.

The expressions for $I_e$ and $I_\epsilon$ can be simplified using the symmetries of the scattering matrix. In fact,  $S^{ee}_{\alpha,\beta}(-E)=\left(S^{ee}_{\alpha,\beta}(E)\right)^*$, and as a consequence:
\be
\label{eq:simmetria-dispari}
1-|S_{\alpha,\beta}^{ee}(-E)|^2 -|S_{\omega,\alpha,\beta}^{ee} (-E)|^2 -|S_{-\omega,\alpha,\beta}^{ee} (-E)|^2  = 1-|S_{\alpha,\beta}^{ee}(E)|^2 -|S_{\omega,\alpha,\beta}^{ee} (E)|^2 -|S_{-\omega,\alpha,\beta}^{ee}(E)|^2 \nonumber  \\
T^{ee}_{\alpha,\beta}(-E) \equiv |S_{-\omega, \alpha,\beta}^{ee} (-E)|^2 - |S_{\omega,\alpha,\beta}^{ee} (-E)|^2 = - |S_{-\omega, \alpha,\beta}^{ee} (E)|^2 - |S_{\omega,\alpha,\beta}^{ee} (E)|^2 = - T^{ee}_{\alpha,\beta}(E) \nonumber \\
T^{eh}_{\alpha,\beta}(-E) \equiv|S_{-\omega, \alpha,\beta}^{eh} (-E)|^2 - |S_{\omega,\alpha,\beta}^{eh} (-E)|^2 = - |S_{-\omega, \alpha,\beta}^{eh} (E)|^2 - |S_{\omega,\alpha,\beta}^{eh} (E)|^2 = - T^{eh}_{\alpha,\beta}(E).
\ee
Using the parity properties of the fermi function, we can therefore rewrite the heat and charge currents as integrals over positive energies only 
\be
\label{eq:energia-positiva}
I_{e,\alpha} &=& 2 \int_0^\infty dE \, \left[ \left( f_0(E+\mu) -f_0(E-\mu ) \right) \left(1-|S_{\alpha,\beta}^{ee}(E)|^2 -|S_{\omega,\alpha,\beta}^{ee}(E)|^2 -|S_{-\omega,\alpha,\beta}^{ee}(E)|^2 \right)  \right] \nonumber \\
& & + \frac{e \omega}{2\pi} \sum_{\beta}  \int_0^\infty dE \, \left[ \frac{\partial f (E-\mu)}{\partial E}\left( T^{ee}_{\alpha,\beta}(E) - T^{eh}_{\alpha,\beta}(E) \right) + \frac{\partial f (E+\mu)}{\partial E}\left( T^{eh}_{\alpha,\beta}(E) - T^{ee}_{\alpha,\beta}(E) \right)   \right],\\
I_{\epsilon,\alpha}  & = & -\frac{\mu}{e} I_{e,\alpha} +  \frac{\omega}{2\pi} \sum_{\beta}  \int_0^\infty dE \, E \,\left[ \frac{\partial f (E-\mu)}{\partial E}\left( T^{ee}_{\alpha,\beta}(E) + T^{eh}_{\alpha,\beta}(E) \right) + \frac{\partial f (E+\mu)}{\partial E}\left( T^{eh}_{\alpha,\beta}(E) + T^{ee}_{\alpha,\beta}(E) \right)   \right].
\ee
Note that at $\mu=0$ $I_{e,\alpha}=0$ identically. This is a consequence of the symmetry of the scattering matrix, and is in contrast with usual results for which the scattering matrix is approximated to be independent of energy. In that case the expressions in \eqref{eq:simmetria-dispari} are even under energy reversal and one has a vanishing heat current as opposed to a finite charge current.

The expressions can be further simplified by computing explicitly 
\be
\sum_{\beta=L,R} T^{ee}_{L, \beta} =  \frac{1}{4}  \delta X_1 \delta X_2{\rm Im} \left[ \frac{\partial S_{L-,L-}}{\partial X_1}\frac{\partial S_{L-,L-}^*}{\partial X_2} + \frac{\partial S_{L-,R-}}{\partial X_1}\frac{\partial S_{L-,R-}^*}{\partial X_2}\right]  = \sum_{\beta=L,R} T^{eh}_{L, \beta},
\ee 
and similarly for $T^{ee}_{R,\beta}$.
This leads to the expressions
\be
I_{e,\alpha}  &=& 2 \int_0^\infty dE \, \left[ \left( f_0(E+\mu) -f_0(E-\mu ) \right) \left(1-|S_{\alpha,\beta}^{ee}(E)|^2 -|S_{\omega,\alpha,\beta}^{ee}(E)|^2 -|S_{-\omega,\alpha,\beta}^{ee}(E)|^2 \right)  \right] , \\
I_{\epsilon,\alpha}  & = & -\frac{\mu}{e} I_{e,\alpha} +   \frac{\omega}{\pi} \sum_{\beta}  \int_0^\infty dE \, E \,\left[ \frac{\partial f (E-\mu)}{\partial E} T^{ee}_{\alpha,\beta}(E) + \frac{\partial f (E+\mu)}{\partial E} T^{ee}_{\alpha,\beta}(E)    \right],
\ee
which appear in the manuscript.
Note that the only time dependence enters through $X_1$ and $X_2$, and one can perform the time integral over the pumping cycle inside the energy integral.
In particular we consider the combination $\mathcal{Q}_\alpha (\mu, T) = \int_0^\tau dt \, (I_{\epsilon,\alpha} + \mu I_{e,\alpha}/e ) $, so that
\be 
\mathcal{Q}_\alpha (\mu, T) = \int_0^\tau dt \, (I_{\epsilon,\alpha} +\frac{\mu}{e}  I_{e,\alpha} ) =
\int_0^\infty dE\, E \, \mathcal{Q}_{0,\alpha} (E) \left[ \frac{\partial f (E-\mu)}{\partial E} + \frac{\partial f (E+\mu)}{\partial E}   \right].
\ee
with
\be
\mathcal{Q}_{0,\alpha}  (E)= 
\frac{1}{\pi}\int_A d X_1 dX_2\sum_{\beta =L,R,\nu=e,h}  {\rm Im} \left[ \frac{\partial S^{e\nu}_{\alpha,\beta}}{\partial X_1}\frac{\partial {S^{e\nu}_{\alpha,\beta}}^*}{\partial X_2}\right]= \int_0^\tau dt \left( \frac{d n(\alpha)}{dX_1}\frac{dX_1}{dt} + \frac{d n(\alpha)}{dX_2}\frac{dX_2}{dt} \right)
\ee
and 
\be
\frac{d n(\alpha)}{dX} = \frac{1}{2\pi}\sum_{\beta=L,R,\nu=e,h} {\rm Im}   \frac{\partial S^{e\nu}_{\alpha,\beta}}{\partial X} {S^{e\nu}_{\alpha,\beta}}^* 
\ee

Note that at $T=0$  $\mathcal{Q}_L (\mu,0)= \mathcal{Q}(\mu)$, which, for $\mu \to 0$ approaches the universal fraction of the solid angle discussed in the manuscript.

\end{document}